\documentclass[runningheads,envcountsame]{llncs}
\usepackage[T1]{fontenc}
\usepackage{graphicx}
\usepackage{hyperref}
\usepackage{color}

\usepackage{amsmath}
\usepackage{amssymb}
\usepackage{catchfile}

\begin{document}

% Abbreviations

% Calligraphical font

\def\AA{{\mathcal A}}
\def\BB{{\mathcal B}}
\def\CC{{\mathcal C}}
\def\DD{{\mathcal D}}
\def\EE{{\mathcal E}}
\def\FF{{\mathcal F}}
\def\GG{{\mathcal G}}
\def\HH{{\mathcal H}}
\def\II{{\mathcal I}}
\def\JJ{{\mathcal J}}
\def\KK{{\mathcal K}}
\def\LL{{\mathcal L}}
\def\MM{{\mathcal M}}
\def\NN{{\mathcal N}}
\def\OO{{\mathcal O}}
\def\PP{{\mathcal P}}
\def\QQ{{\mathcal Q}}
\def\RR{{\mathcal R}}
\def\SS{{\mathcal S}}
\def\TT{{\mathcal T}}
\def\UU{{\mathcal U}}
\def\VV{{\mathcal V}}
\def\WW{{\mathcal W}}
\def\XX{{\mathcal X}}
\def\YY{{\mathcal Y}}
\def\ZZ{{\mathcal Z}}

% Bold font

\def\bA{{\mathbf A}}
\def\bB{{\mathbf B}}
\def\bC{{\mathbf C}}
\def\bD{{\mathbf D}}
\def\bE{{\mathbf E}}
\def\bF{{\mathbf F}}
\def\bG{{\mathbf G}}
\def\bH{{\mathbf H}}
\def\bI{{\mathbf I}}
\def\bJ{{\mathbf J}}
\def\bK{{\mathbf K}}
\def\bL{{\mathbf L}}
\def\bM{{\mathbf M}}
\def\bN{{\mathbf N}}
\def\bO{{\mathbf O}}
\def\bP{{\mathbf P}}
\def\bQ{{\mathbf Q}}
\def\bR{{\mathbf R}}
\def\bS{{\mathbf S}}
\def\bT{{\mathbf T}}
\def\bU{{\mathbf U}}
\def\bV{{\mathbf V}}
\def\bW{{\mathbf W}}
\def\bX{{\mathbf X}}
\def\bY{{\mathbf Y}}
\def\bZ{{\mathbf Z}}

% Blackboard font

\def\IB{{\mathbb{B}}}
\def\IC{{\mathbb{C}}}
\def\IF{{\mathbb{F}}}
\def\IN{{\mathbb{N}}}
\def\IP{{\mathbb{P}}}
\def\IQ{{\mathbb{Q}}}
\def\IR{{\mathbb{R}}}
\def\IS{{\mathbb{S}}}
\def\IT{{\mathbb{T}}}
\def\IZ{{\mathbb{Z}}}

\def\IIB{{\mathbb{\mathbf B}}}
\def\IIC{{\mathbb{\mathbf C}}}
\def\IIN{{\mathbb{\mathbf N}}}
\def\IIQ{{\mathbb{\mathbf Q}}}
\def\IIR{{\mathbb{\mathbf R}}}
\def\IIZ{{\mathbb{\mathbf Z}}}

% Definitions

%\def\IF{\quad\mbox{if}\quad}
\def\ELSE{\quad\mbox{else}\quad}
\def\WITH{\quad\mbox{with}\quad}
\def\FOR{\quad\mbox{for}\quad}
\def\AND{\;\mbox{and}\;}
\def\OR{\;\mbox{or}\;}

\def\To{\longrightarrow}
\def\TO{\Longrightarrow}
\def\In{\subseteq}
\def\sm{\setminus}
\def\Inneq{\In_{\!\!\!\!/}}
\def\dmin{\mathop{\dot{-}}}
\def\splus{\oplus}
\def\SEQ{\triangle}
\def\DIV{\uparrow}
\def\INV{\leftrightarrow}
\def\SET{\Diamond}

\def\kto{\equiv\!\equiv\!>}
\def\kin{\subset\!\subset}
\def\pto{\leadsto}
\def\into{\hookrightarrow}
\def\onto{\to\!\!\!\!\!\to}
\def\prefix{\sqsubseteq}
\def\rel{\leftrightarrow}
\def\mto{\rightrightarrows}

\def\B{{\mathsf{{B}}}}
\def\D{{\mathsf{{D}}}}
\def\G{{\mathsf{{G}}}}
\def\E{{\mathsf{{E}}}}
\def\J{{\mathsf{{J}}}}
\def\K{{\mathsf{{K}}}}
\def\L{{\mathsf{{L}}}}
\def\R{{\mathsf{{R}}}}
\def\T{{\mathsf{{T}}}}
\def\U{{\mathsf{{U}}}}
\def\W{{\mathsf{{W}}}}
\def\Z{{\mathsf{{Z}}}}
\def\w{{\mathsf{{w}}}}
\def\HP{{\mathsf{{H}}}}
\def\C{{\mathsf{{C}}}}
\def\Tot{{\mathsf{{Tot}}}}
\def\Fin{{\mathsf{{Fin}}}}
\def\Cof{{\mathsf{{Cof}}}}
\def\Cor{{\mathsf{{Cor}}}}
\def\Equ{{\mathsf{{Equ}}}}
\def\Com{{\mathsf{{Com}}}}
\def\Inf{{\mathsf{{Inf}}}}

\def\Tr{{\mathrm{Tr}}}
\def\Sierp{{\mathrm Sierpi{\'n}ski}}
\def\psisierp{{\psi^{\mbox{\scriptsize\Sierp}}}}
\def\cl{{\mathrm{{cl}}}}
\def\Haus{{\mathrm{{H}}}}
\def\Ls{{\mathrm{{Ls}}}}
\def\Li{{\mathrm{{Li}}}}

\def\CL{\mathsf{CL}}
\def\ACC{\mathsf{ACC}}
\def\DNC{\mathsf{DNC}}
\def\ATR{\mathsf{ATR}}
\def\LPO{\mathsf{LPO}}
\def\LLPO{\mathsf{LLPO}}
\def\WKL{\mathsf{WKL}}
\def\RCA{\mathsf{RCA}}
\def\ACA{\mathsf{ACA}}
\def\SEP{\mathsf{SEP}}
\def\BCT{\mathsf{BCT}}
\def\IVT{\mathsf{IVT}}
\def\IMT{\mathsf{IMT}}
\def\OMT{\mathsf{OMT}}
\def\CGT{\mathsf{CGT}}
\def\UBT{\mathsf{UBT}}
\def\BWT{\mathsf{BWT}}
\def\HBT{\mathsf{HBT}}
\def\BFT{\mathsf{BFT}}
\def\FPT{\mathsf{FPT}}
\def\WAT{\mathsf{WAT}}
\def\LIN{\mathsf{LIN}}
\def\B{\mathsf{B}}
\def\BF{\mathsf{B_\mathsf{F}}}
\def\BI{\mathsf{B_\mathsf{I}}}
\def\C{\mathsf{C}}
\def\CF{\mathsf{C_\mathsf{F}}}
\def\CN{\mathsf{C_{\IN}}}
\def\CI{\mathsf{C_\mathsf{I}}}
\def\CK{\mathsf{C_\mathsf{K}}}
\def\CA{\mathsf{C_\mathsf{A}}}
\def\WPO{\mathsf{WPO}}
\def\WLPO{\mathsf{WLPO}}
\def\MP{\mathsf{MP}}
\def\BD{\mathsf{BD}}
\def\Fix{\mathsf{Fix}}
\def\Mod{\mathsf{Mod}}

\def\s{\mathrm{s}}
\def\r{\mathrm{r}}
\def\w{\mathsf{w}}

\def\leqm{\mathop{\leq_{\mathrm{m}}}}
\def\equivm{\mathop{\equiv_{\mathrm{m}}}}
\def\leqT{\mathop{\leq_{\mathrm{T}}}}
\def\lT{\mathop{<_{\mathrm{T}}}}
\def\nleqT{\mathop{\not\leq_{\mathrm{T}}}}
\def\equivT{\mathop{\equiv_{\mathrm{T}}}}
\def\nequivT{\mathop{\not\equiv_{\mathrm{T}}}}
\def\leqwtt{\mathop{\leq_{\mathrm{wtt}}}}
\def\equiPT{\mathop{\equiv_{\P\mathrm{T}}}}
\def\leqW{\mathop{\leq_{\mathrm{W}}}}
\def\equivW{\mathop{\equiv_{\mathrm{W}}}}
\def\leqtW{\mathop{\leq_{\mathrm{tW}}}}
\def\leqSW{\mathop{\leq_{\mathrm{sW}}}}
\def\equivSW{\mathop{\equiv_{\mathrm{sW}}}}
\def\leqPW{\mathop{\leq_{\widehat{\mathrm{W}}}}}
\def\equivPW{\mathop{\equiv_{\widehat{\mathrm{W}}}}}
\def\leqFPW{\mathop{\leq_{\mathrm{W}^*}}}
\def\equivFPW{\mathop{\equiv_{\mathrm{W}^*}}}
\def\leqWW{\mathop{\leq_{\overline{\mathrm{W}}}}}
\def\nleqW{\mathop{\not\leq_{\mathrm{W}}}}
\def\nleqSW{\mathop{\not\leq_{\mathrm{sW}}}}
\def\lW{\mathop{<_{\mathrm{W}}}}
\def\lSW{\mathop{<_{\mathrm{sW}}}}
\def\nW{\mathop{|_{\mathrm{W}}}}
\def\nSW{\mathop{|_{\mathrm{sW}}}}
\def\leqt{\mathop{\leq_{\mathrm{t}}}}
\def\equivt{\mathop{\equiv_{\mathrm{t}}}}
\def\leqtop{\mathop{\leq_\mathrm{t}}}
\def\equivtop{\mathop{\equiv_\mathrm{t}}}

\def\bigtimes{\mathop{\mathsf{X}}}

\def\leqm{\mathop{\leq_{\mathrm{m}}}}
\def\equivm{\mathop{\equiv_{\mathrm{m}}}}
\def\leqT{\mathop{\leq_{\mathrm{T}}}}
\def\leqM{\mathop{\leq_{\mathrm{M}}}}
\def\equivT{\mathop{\equiv_{\mathrm{T}}}}
\def\equiPT{\mathop{\equiv_{\P\mathrm{T}}}}
\def\leqW{\mathop{\leq_{\mathrm{W}}}}
\def\equivW{\mathop{\equiv_{\mathrm{W}}}}
\def\nequivW{\mathop{\not\equiv_{\mathrm{W}}}}
\def\leqSW{\mathop{\leq_{\mathrm{sW}}}}
\def\equivSW{\mathop{\equiv_{\mathrm{sW}}}}
\def\leqPW{\mathop{\leq_{\widehat{\mathrm{W}}}}}
\def\equivPW{\mathop{\equiv_{\widehat{\mathrm{W}}}}}
\def\nleqW{\mathop{\not\leq_{\mathrm{W}}}}
\def\nleqSW{\mathop{\not\leq_{\mathrm{sW}}}}
\def\lW{\mathop{<_{\mathrm{W}}}}
\def\lSW{\mathop{<_{\mathrm{sW}}}}
\def\nW{\mathop{|_{\mathrm{W}}}}
\def\nSW{\mathop{|_{\mathrm{sW}}}}

\def\botW{\mathbf{0}}
\def\midW{\mathbf{1}}
\def\topW{\mathbf{\infty}}

\def\pol{{\leq_{\mathrm{pol}}}}
\def\rem{{\mathop{\mathrm{rm}}}}

\def\cc{{\mathrm{c}}}
\def\d{{\,\mathrm{d}}}
\def\e{{\mathrm{e}}}
\def\ii{{\mathrm{i}}}

\def\Cf{C\!f}
\def\id{{\mathrm{id}}}
\def\pr{{\mathrm{pr}}}
\def\inj{{\mathrm{inj}}}
\def\cf{{\mathrm{cf}}}
\def\dom{{\mathrm{dom}}}
\def\range{{\mathrm{range}}}
\def\graph{{\mathrm{graph}}}
\def\Graph{{\mathrm{Graph}}}
\def\epi{{\mathrm{epi}}}
\def\hypo{{\mathrm{hypo}}}
\def\Lim{{\mathrm{Lim}}}
\def\diam{{\mathrm{diam}}}
\def\dist{{\mathrm{dist}}}
\def\supp{{\mathrm{supp}}}
\def\union{{\mathrm{union}}}
\def\fiber{{\mathrm{fiber}}}
\def\ev{{\mathrm{ev}}}
\def\mod{{\mathrm{mod}}}
\def\sat{{\mathrm{sat}}}
\def\seq{{\mathrm{seq}}}
\def\lev{{\mathrm{lev}}}
\def\mind{{\mathrm{mind}}}
\def\arccot{{\mathrm{arccot}}}
\def\cl{{\mathrm{cl}}}

\def\Add{{\mathrm{Add}}}
\def\Mul{{\mathrm{Mul}}}
\def\SMul{{\mathrm{SMul}}}
\def\Neg{{\mathrm{Neg}}}
\def\Inv{{\mathrm{Inv}}}
\def\Ord{{\mathrm{Ord}}}
\def\Sqrt{{\mathrm{Sqrt}}}
\def\Re{{\mathrm{Re}}}
\def\Im{{\mathrm{Im}}}
\def\Sup{{\mathrm{Sup}}}

\def\LSC{{\mathcal LSC}}
\def\USC{{\mathcal USC}}

\def\CE{{\mathcal{E}}}
\def\Pref{{\mathrm{Pref}}}

\def\Baire{\IN^\IN}

\def\TRUE{{\mathrm{TRUE}}}
\def\FALSE{{\mathrm{FALSE}}}

\def\co{{\mathrm{co}}}

\def\BBB{{\tt B}}

\newcommand{\SO}[1]{{{\mathbf\Sigma}^0_{#1}}}
\newcommand{\SI}[1]{{{\mathbf\Sigma}^1_{#1}}}
\newcommand{\PO}[1]{{{\mathbf\Pi}^0_{#1}}}
\newcommand{\PI}[1]{{{\mathbf\Pi}^1_{#1}}}
\newcommand{\DO}[1]{{{\mathbf\Delta}^0_{#1}}}
\newcommand{\DI}[1]{{{\mathbf\Delta}^1_{#1}}}
\newcommand{\sO}[1]{{\Sigma^0_{#1}}}
\newcommand{\sI}[1]{{\Sigma^1_{#1}}}
\newcommand{\pO}[1]{{\Pi^0_{#1}}}
\newcommand{\pI}[1]{{\Pi^1_{#1}}}
\newcommand{\dO}[1]{{{\Delta}^0_{#1}}}
\newcommand{\dI}[1]{{{\Delta}^1_{#1}}}
\newcommand{\sP}[1]{{\Sigma^\P_{#1}}}
\newcommand{\pP}[1]{{\Pi^\P_{#1}}}
\newcommand{\dP}[1]{{{\Delta}^\P_{#1}}}
\newcommand{\sE}[1]{{\Sigma^{-1}_{#1}}}
\newcommand{\pE}[1]{{\Pi^{-1}_{#1}}}
\newcommand{\dE}[1]{{\Delta^{-1}_{#1}}}

\newcommand{\dBar}[1]{{\overline{\overline{#1}}}}

\def\QED{$\hspace*{\fill}\Box$}
\def\rand#1{\marginpar{\rule[-#1 mm]{1mm}{#1mm}}}

\def\BL{\BB}

% Commands

\newcommand{\bra}[1]{\langle#1|}
\newcommand{\ket}[1]{|#1\rangle}
\newcommand{\braket}[2]{\langle#1|#2\rangle}

\newcommand{\ind}[1]{{\em #1}\index{#1}}
\newcommand{\mathbox}[1]{\[\fbox{\rule[-4mm]{0cm}{1cm}$\quad#1$\quad}\]}

% Environments

%\newenvironment{proof}{{\mathbf Proof.}\begin{small}}{\end{small}\QED\\}
%\newenvironment{cases}{\left\{\begin{array}{ll}}{\end{array}\right.}
\newenvironment{eqcase}{\left\{\begin{array}{lcl}}{\end{array}\right.}

\def\IVP{\mathsf{IVP}}
\def\sign{\mathrm{sign}}

\title{Computability of Initial Value Problems}
%
%\titlerunning{Abbreviated paper title}
% If the paper title is too long for the running head, you can set
% an abbreviated paper title here
%
\author{Vasco Brattka\inst{1,2}\orcidID{0000-0003-4664-2183} \and\\
Hendrik Smischliaew\inst{1}\orcidID{0009-0009-5917-3452}}

\authorrunning{V.\ Brattka and H.\ Smischliaew}
% First names are abbreviated in the running head.
% If there are more than two authors, 'et al.' is used.
%
\institute{Fakult\"at f\"ur Informatik, Universit\"at der Bundeswehr M\"unchen, Werner-Heisenberg-Weg 39, 85577 Neubiberg, Germany \and
Department of Mathematics and Applied Mathematics, University of Cape Town, Private Bag X3, Rondebosch 7701, South Africa\\
\email{Vasco.Brattka@cca-net.de}\\
\email{hendrik@hsmischliaew.de}}
\maketitle              % typeset the header of the contribution
\begin{abstract}
We demonstrate that techniques of Weihrauch complexity can be used to get easy and elegant proofs
of known and new results on initial value problems.
Our main result is that solving continuous initial value problems is Weihrauch equivalent to weak K\H{o}nig's lemma, 
even if only solutions with maximal domains of existence are considered. 
This result simultaneously generalizes negative and positive results by Aberth and by Collins and Gra\c{c}a, respectively.
It can also be seen as a uniform version of a Theorem of Simpson.
Beyond known techniques we exploit for the proof that weak K\H{o}nig's lemma is closed under infinite loops. 
One corollary of our main result is that solutions with maximal domain of existence
of continuous initial value problems can be computed non-deterministically,
and for computable instances there are always solutions that are low as points in the function space.
Another corollary is that in the case that there is a fixed finite number of solutions, 
these solutions are all computable for computable instances and they can be found
uniformly in a finite mind-change computation.

\keywords{Computable analysis  \and Weihrauch complexity \and ordinary differential equations.}
\end{abstract}
\section{Introduction}

We consider {\em initial value problems} of the form
\begin{eqnarray}
\left\{\begin{array}{ll}
y'(x)=f(x,y(x))\\
y'(x_0)=y_0
\end{array}\right.
\label{eq:IVP}
\end{eqnarray}
for continuous functions $f:U\to\IR^n$ with $U\In\IR\times\IR^{n}$ and $(x_0,y_0)\in U$.
A {\em solution} of such a problem is a differentiable function $y:I\to\IR^n$ that satisfies the equations in~(\ref{eq:IVP})
on some interval $I\In\IR$ with $x_0\in I$.
Being a solution entails that $(x,y(x))\in U$ for all $x\in I$. 
Any such solution $y$ is automatically continuously differentiable.
We say that $I=(a,b)$ with $a\in\IR\cup\{-\infty\}$ and $b\in\IR\cup\{\infty\}$ is a 
{\em maximal interval of existence} for a solution $y:I\to\IR^n$, if $y$ is a solution and no proper
extension of $y$ to a strictly larger interval of the above form is a solution.

There are two classical theorems that guarantee the existence of solutions of initial value problems, which are relevant in our
context. The Picard-Lindel\"of theorem guarantees the uniqueness of the solution on some 
small interval, in the case that $f$ satisfies some Lipschitz condition~\cite[Theorem~2.2]{Tes12}.

\begin{theorem}[Picard-Lindel\"of]
\label{thm:Picard}
Let $U\In\IR\times\IR^n$ be open with $(x_0,y_0)\in U$ 
and let $f:U\to\IR^n,(t,s)\mapsto f(t,s)$ be continuous 
and locally Lipschitz continuous in the second argument $s\in\IR^n$, uniformly with respect to the first argument $t\in\IR$.
Then the initial value problem (\ref{eq:IVP}) has a unique solution $y:[x_0-\varepsilon,x_0+\varepsilon]\to\IR^n$
on some interval with $\varepsilon>0$.
\end{theorem}

Here local Lipschitz continuity in the second argument, uniformly with respect to the first argument, 
means that for any compact subset $K\In U$
there is a Lipschitz constant in the second argument that works uniformly for each fixed first argument (see~\cite{Tes12}).
The theorem can be proved with the help of the Banach fixed-point theorem, applied to a {\em Picard operator}
$T:D\to D$ (see~(\ref{eqn:Picard}) below)
with a suitably chosen domain $D\In\CC([a,b],\IR^n)$ of continuous functions.
It is not too difficult to see that the fixed points of $T$ correspond to the solutions of 
(\ref{eq:IVP}) for the respective interval $I=[a,b]$ (see~\cite[Section~2.2]{Tes12}).

In the case that $f$ is only continuous and not necessarily Lipschitz continuous,
it is still guaranteed that there are solutions, but not necessarily a unique one.
This existence of solutions follows from the Peano theorem~\cite[Theorem~2.19]{Tes12}.

\begin{theorem}[Peano]
\label{thm:Peano}
Let $U\In\IR\times\IR^n$ be open with $(x_0,y_0)\in U$ 
and let $f:U\to\IR^n$ be continuous.
Then the initial value problem (\ref{eq:IVP}) has a solution $y:[x_0-\varepsilon,x_0+\varepsilon]\to\IR^n$
on some interval with $\varepsilon>0$.
\end{theorem}

Again this result can be proved with the help of a suitable Picard operator as in (\ref{eqn:Picard}) below, 
but in this case one needs to apply a different fixed-point theorem, such as the Schauder fixed-point theorem
in order to obtain a fixed point. This explains why this result is less constructive
than the Picard-Lindel\"of theorem.

The study of initial value problems has a long tradition in computable analysis (for a survey see
Gra\c{c}a and Zhong~\cite{GZ21}). One of the earliest results is by Aberth~\cite{Abe71}, who
proved that even for computable $f$ there is not necessarily a computable solution.

\begin{theorem}[Aberth]
\label{thm:Aberth}
There exists a computable function $f:[-1,1]^2\to\IR$ such that the initial value problem (\ref{eq:IVP})
with $x_0=y_0=0$ has no computable solution $y:I\to\IR$, defined on some interval $I\In\IR$ with interior point $x_0=0$.
\end{theorem}

This example was later strengthened by Pour-El and Richards~\cite{PR79}, who proved that
there is even such a counterexample that has no computable solution irrespectively of the chosen initial value.

On the positive side, it follows from a computable version of the Theorem of Picard-Lindel\"of that
for computable $f$ that satisfies some suitable Lipschitz condition
there is a unique computable solution of the initial value problem~(\ref{eq:IVP}) for some interval. 
And more than this, even the solution on 
the  maximal interval of existence $I=(a,b)$ is computable by a theorem of Gra\c{c}a, Zhong and Buescu~\cite[Theorem~3.1]{GZB09}.
Ruohonen~\cite{Ruo96} showed that in general, if the solution is uniquely determined for computable $f$,
then the solution is computable. 
This result was extended to maximal domains of existence by Collins and Gra\c{c}a~\cite[Theorem~21]{CG09}.

\begin{theorem}[Collins and Gra\c{c}a]
\label{thm:Collins}
Let $f:U\to\IR^n$ be a computable function on a c.e.\ open set $U\In\IR\times\IR^n$ and let $(x_0,y_0)\in U$ be computable. 
Suppose that (\ref{eq:IVP}) has a unique solution $y:I\to\IR^n$ with a maximal interval $I=(a,b)\In\IR$ of existence
such that $x_0\in I$. Then $I$ is c.e.\ open and $y$ is computable.
\end{theorem}

More than this, the authors have shown that given $(f,x_0,y_0)$, one can even uniformly compute $(y, I)$.

In a seemingly different direction it has been proved in reverse mathematics by Simpson that the
Peano existence theorem is equivalent to weak K\H{o}nig's lemma over the base system of
recursive comprehension $\RCA_0$~\cite{Sim84} and \cite[Theorem IV.8.1]{Sim09}.

\begin{theorem}[Simpson]
\label{thm:Simpson}
In second-order arithmetic (a suitable version of) 
the theorem of Peano is equivalent to weak K\H{o}nig's lemma $\WKL_0$
over $\RCA_0$.
\end{theorem}

One direction of the proof of this result is essentially based on the proof idea of Aberth's theorem (Theorem~\ref{thm:Aberth}),
whereas the other direction uses an appropriate version of the Schauder fixed point theorem.

Our goal here is to establish a similar result for Weihrauch complexity, which offers a computational
way of classifying the computational content of mathematical problems (see \cite{BGP21} for a recent survey).
Our proof incorporates ideas of Aberth, Simpson, and of Gra\c{c}a, Zhong and Buescu~\cite{GZB09},
but it also requires some new techniques and ideas, for instance regarding infinite loops.
Our main result is the following Weihrauch complexity classification of the initial value problem.

\begin{theorem}
\label{thm:main}
The following are pairwise (strongly) Weihrauch equivalent:
\begin{enumerate}
\item The initial value problem $\IVP$ for the special case $x_0=y_0=0$ and $n=2$.
\item The initial value problem $\IVP$.
\item The initial value problem for maximal domains of existence $\IVP_{\max}$.
\item Weak K\H{o}nig's lemma $\WKL$.
\end{enumerate}
\end{theorem}

Here $\IVP$ denotes the problem, given $(f,U,x_0,y_0)\in\CC(U,\IR^n)\times\OO(\IR^{n+1})\times\IR^{n+1}$, 
find $(y,I)\in\CC(I,\IR^n)\times\OO(\IR)$ such that $y:I\to\IR^n$ is a solution of (\ref{eq:IVP}).
Here $\CC(X,Y)$ denotes the space of continuous functions $f:X\to Y$ and $\OO(X)$ the space
of open subsets $U\In X$, both represented in the standard way. 
The problem $\IVP_{\max}$ is defined analogously, except that $I$ is additionally required
to be a maximal domain of existence. 
And as usual, $\WKL$ denotes the problem, given an infinite binary tree $T$, find an infinite path $p\in[T]$
of $T$.

The equivalence of 2.\ and 4.\ in Theorem~\ref{thm:main} can be seen as a uniform 
version of the theorem of Simpson (Theorem~\ref{thm:Simpson}). However, we also obtain a version of the theorems
of Aberth (Theorem~\ref{thm:Aberth}) and of Collins and Gra\c{c}a (Theorem~\ref{thm:Collins})
as immediate corollaries of Theorem~\ref{thm:main}.
And more than this, we can say something on solutions with maximal domain of existence
in the general case.

\begin{corollary}
\label{cor:main}
Let $f:U\to\IR^n$ be computable with a c.e.\ open set $U\In\IR\times\IR^n$ and let $(x_0,y_0)\in U$ be computable.
Then there exists a solution $y:I\to\IR^n$ of (\ref{eq:IVP}) with a maximal domain $I=(a,b)\In\IR$ of existence
such that $y$ is low as a point in $\CC(I,\IR^n)$. And given $(f,U,x_0,y_0)$, a solution $(y,I)$ can be found
uniformly in a non-deterministic way.
\end{corollary}

These statements are immediate consequences of Theorem~\ref{thm:main}, as it
is a property of the Weihrauch equivalence class of $\WKL$ that all problems in this
class are non-deterministically computable and computable instances of such problems
have low solutions \cite[Corollary~7.13, Theorem~8.3]{BBP12}.
The fact that we obtain Theorem~\ref{thm:Collins} as a corollary also exploits
the fact that single-valued functions below $\WKL$ are automatically computable~\cite[Corollary~5.2]{BBP12}.
Using further well-known results from Weihrauch complexity~\cite[Proposition~3.2]{LRP15a}
we can, for instance, obtain the following result, which is reminiscent of results by Hauck~\cite{Hau85}.

\begin{corollary}
\label{cor:finite}
Let $f:U\to\IR^n$ be computable with a c.e.\ open set $U\In\IR\times\IR^n$ and computable $(x_0,y_0)\in U$.
If there are only finitely many solutions $y:I\to\IR^n$ of (\ref{eq:IVP}) with a maximal domain $I=(a,b)\In\IR$ of existence,
then all of these solutions $y$ are computable, and all the maximal domains of existence are c.e.\ open.
Given $(f,U,x_0,y_0)$ with a fixed finite number of solutions, one such maximal solution $(y,I)$ can be found
uniformly with a finite mind-change computation.
\end{corollary}

We close this section with a description of the further content of this article.
In the next section we provide some basic definitions that are required to define
the initial value problem formally, and we introduce some concepts from Weihrauch complexity.
In Section~\ref{sec:IVP-WKL} we prove that the initial value problem 
is Weihrauch reducible to weak K\H{o}nig's lemma. 
In Section~\ref{sec:IVP-MAX-WKL} we strengthen this result to the initial value problem with maximal
domain of existence.
In Section~\ref{sec:WKL-IVP} we discuss the reduction in the opposite direction.\footnote{The results presented in this article are based on the master's thesis of the second author~\cite{Smi24}, which was written under supervision of the first author.}

\section{Weihrauch complexity and the initial value problem}
\label{sec:Weihrauch}

We introduce some concepts from computable analysis and Weihrauch complexity
and we refer the reader to \cite{BH21,Wei00} for all concepts that have not been introduced here.
We follow the representation based approach to computable analysis and we recall
that a {\em representation} of a space $X$ is a surjective partial map $\delta_X:\In\IN^\IN\to X$.
In this case $(X,\delta_X)$ is called a {\em represented space}. If we have two represented spaces
$(X,\delta_X)$ and $(Y,\delta_Y)$, then we automatically have a representation $\delta_{\CC(X,Y)}$ of the space
of functions $f:X\to Y$ that have a continuous realizer. Here $F:\In\IN^\IN\to\IN^\IN$
is called a {\em realizer} of some partial multivalued function $f:\In X\mto Y$, if
\[\delta_YF(p)\in f\delta_X(p)\]
for all $p\in\dom(f\delta_X)$. In this situation we also write $F\vdash f$.
It is well-known that there are universal functions $\U:\In\IN^\IN\to\IN^\IN$ such that
for every continuous $F:\In\IN^\IN\to\IN^\IN$ there is some $q\in\IN^\IN$ such that
$F(p)=\U\langle q,p\rangle$ for all $p\in\dom(F)$. Here $\langle.\rangle$ denotes some
standard pairing function on Baire space $\IN^\IN$ (we use this notation for pairs as well as for the pairing of sequences).
For short we write $\U_q(p):=\U\langle q,p\rangle$ for all $q,p\in\IN^\IN$.
Now we obtain a representation
$\delta_{\CC(X,Y)}$ of the set $\CC(X,Y)$ of total singlevalued functions $f:X\to Y$ with continuous realizers by
\[\delta_{\CC(X,Y)}(q)=f:\iff\U_q\vdash f.\]
It is well-known that for admissibly represented $T_0$--spaces $X,Y$ the function space $\CC(X,Y)$ consists
exactly of the usual continuous functions (see \cite{BH21,Wei00} for more details).
The first difficulty that we face is that we need representations of function spaces $\CC(U,Y)$ for
varying domains $U$. Such representations have not been widely used in the literature
and we use coproducts of the following form for this purpose.

\begin{definition}[Coproduct function spaces]
\rm
Let $X,Y$ be represented spaces and let
$(\PP(X),\delta_\PP)$ be a represented space with $\PP(X)\In 2^X$.
Then 
\begin{eqnarray*}
&\bigsqcup_{A\in\PP(X)}\CC(A,Y):=\{(f,A):A\in\PP(X)\mbox{ and }f\in\CC(A,Y)\}\
\end{eqnarray*}
denotes the {\em coproduct function space} that we represent by $\delta$,
defined by
\[\delta\langle q,p\rangle=(f,A):\iff \delta_\PP(p)=A\mbox{ and }\U_q\vdash f\]
for all total singlevalued functions $f:A\to Y$ in $\CC(A,Y)$ with $A\in\PP(X)$.
\end{definition}

We note that the representations of $X$ and $Y$ occur implicitly in the definition of $\vdash$.
Typically, we will use for $\PP(X)$ the set $\OO(X)$ of open subsets of $X$. 
This set is represented via
characteristic functions in $\CC(X,\IS)$ to {\em Sierpi\'nski space} $\IS=\{0,1\}$,
which in turn is represented in the standard way. Also the {\em Euclidean space} $\IR^n$
is represented in the standard way. The space of closed sets $\AA_-(X)$ equipped with negative
information is represented using complements of open sets in $\OO(X)$. The space of compact
subsets $\KK_-(X)$ (of some computable metric space $X$) is represented via the 
{\em universal map}
\[\forall_K:\OO(X)\to\IS, \forall_K(U)=1:\iff K\In U,\] 
see~\cite{Pau16,Sch21} for details. The computable points in $\OO(X)$, $\AA_-(X)$ and $\KK_-(X)$
are called {\em c.e.\ open}, {\em co-c.e.\ closed} and {\em co-c.e.\ compact} sets, respectively. 

In general a {\em problem} is a multivalued function $f:\In X\mto Y$ on represented
spaces $X,Y$ that has a realizer. A typical example of a problem is {\em weak K\H{o}nig's lemma}
\[\WKL:\In\Tr\mto2^\IN,T\mapsto[T]\]
that is defined for all infinite binary trees $T\in\Tr$ and maps those to the set $[T]$ of infinite paths.
Here $\Tr$ denotes the set of binary trees represented via their characteristic functions.
Hence, input and output space of $\WKL$ can be seen as subspaces of $\IN^\IN$. 
Another typical problem is {\em compact choice}
\[\K_X:\In\KK_-(X)\mto X,K\mapsto K,\]
which maps any non-empty compact set $K\In X$ to its points. 
The problem $\LLPO:=\K_{\{0,1\}}$ is also know as {\em lesser limited problem of omniscience}.
The following was essentially proved in~\cite{GM09} (see also \cite[Theorem~8.5]{BG11}).

\begin{theorem}[Gherardi and Marcone]
\label{thm:K-WKL}
$\K_X\leqSW\WKL$ for every computable metric space $X$.
\end{theorem}

We recall that a {\em computable metric space} is a metric space $(X,d)$ together with
a dense sequence $\alpha:\IN\to X$ such that $d\circ(\alpha\times\alpha):\IN\times\IN\to\IR$ is computable.

We can now define our versions of the initial value problem formally. For the remainder
we assume that $n\geq2$ is some fixed dimension. With a little more effort, we could
also make the dimension variable in the coproduct.

\begin{definition}[Initial value problem]
\rm
By 
\begin{eqnarray*}
&\IVP:\In\bigsqcup_{U\in\OO(\IR^{n+1})}\CC(U,\IR^n)\times\IR\times\IR^{n}\mto\bigsqcup_{I=(a,b)\in\OO(\IR)}\CC(I,\IR^n)
\end{eqnarray*}
we denote the {\em initial value problem}, defined by
\[\IVP(f,U,x_0,y_0):=\{(y,I):I=(a,b)\in\OO(\IR)\mbox{ and }y:I\to\IR^n\mbox{ solves }(\ref{eq:IVP})\}\]
with $\dom(\IVP):=\{(f,U,x_0,y_0):(x_0,y_0)\in U\}$.
By $\IVP_{\max}$ we denote the restriction of $\IVP$ in the image to such solutions ${y:I\to\IR^n}$ for which
$I$ is a maximal domain of existence.
\end{definition}

We use the usual concept of Weihrauch reducibility (see \cite{BGP21} for a survey) in order to compare problems.
Here $\id:\IN^\IN\to\IN^\IN$ denotes the identity on Baire space.

\begin{definition}[Weihrauch reducibility]
\rm
Let $f:\In X\mto Y$ and $g:\In Z\mto W$ be problems. We say that
\begin{enumerate}
\item $f$ is {\em Weihrauch reducible} to $g$, in symbols $f\leqW g$, if there are computable
         $H,K:\In\IN^\IN\to\IN^\IN$ such that $H\langle\id,GK\rangle\vdash f$, whenever $G\vdash g$ holds. 
\item $f$ is {\em strongly Weihrauch reducible} to $g$, in symbols $f\leqSW g$, if there are computable
         $H,K:\In\IN^\IN\to\IN^\IN$ such that $HGK\vdash f$, whenever $G\vdash g$ holds. 
\end{enumerate}
\end{definition}

As usual, we denote the corresponding equivalences by $\equivW$ and $\equivSW$, respectively. 
We also need a number of operators on problems that are commonly used in Weihrauch complexity,
such as the {\em parallelization}
\[\widehat{f}:\In X^\IN\mto Y^\IN,(x_n)_{n\in\IN}\mapsto\bigtimes_{n\in\IN} f(x_n),\]
which is defined for every problem $f:\In X\mto Y$.
We also define the concept of an {\em inverse limit} on problems, which can be 
seen as an infinite loop operation. For technical simplicity we define this for problems on Baire space.

\begin{definition}[Inverse limit]
\rm
Let $f:\In\IN^\IN\mto\IN^\IN$ be a problem. Then we define
the {\em inverse limit} $f^\infty:\In\IN^\IN\mto\IN^\IN$ of $f$
by
\[f^\infty(q_0):=\{\langle q_0,q_1,q_2,...\rangle\in\IN^\IN:(\forall i)\;q_{i+1}\in \U\circ\langle \id\times f\rangle(q_i)\}\]
where $\dom(f^\infty)$ consists of all $q_0\in\IN^\IN$ such that
$A_0:=\{q_0\}\In\dom(\U\circ\langle\id\times f\rangle)$ and $A_{i+1}:=\U\circ\langle \id\times f\rangle(A_i)\In\dom(\U\circ\langle\id\times f\rangle)$ for all $i\in\IN$.
\end{definition}

Using standard techniques we can extend this definition to represented spaces.
The main result we need about the inverse limit is that weak K\H{o}nig's lemma is closed under this operation,
which was proved in~\cite{Bra24}. We also use the fact that the parallelization of $\LLPO$ is equivalent
to weak K\H{o}nig's lemma~\cite[Theorem~8.2]{BG11}.

\begin{proposition}
\label{prop:WKL}
$\WKL^\infty\equivSW\WKL\equivSW\widehat{\LLPO}$.
\end{proposition}

It has also been proved in~\cite{Bra24} that $f\mapsto f^\infty$ is a monotone operation
with respect to (strong) Weihrauch reducibility.
Now we are well prepared to discuss the proof of our main result.

\section{Reduction of $\IVP$ to $\WKL$}
\label{sec:IVP-WKL}

In this section we prove that the initial value problem $\IVP$ is reducible to weak K\H{o}nig's lemma $\WKL$.
The main idea is pretty simple: the solutions of $\IVP$ can be obtained as fixed points of the Picard operator~(\ref{eqn:Picard}) below
for a suitable domain $D$. Most of the work goes into choosing an appropriate compact $D$.
Finding a fixed point can then be achieved with the help of compact choice $\K_X$ for a suitable space $X$ (see below).

We borrow some proof ideas from Simpson~\cite{Sim84} and \cite[Theorem IV.8.1]{Sim09}.
The difference to the proof in reverse mathematics is that we do not need to re-prove the Peano theorem (see Theorem~\ref{thm:Peano}) 
in an effective form, but we can just use the classical theorem that already guarantees the existence of solutions
and hence the existence of fixed points of the Picard operator. Even though computable versions of fixed-point theorems
exist (see~\cite{BLRMP19}), we do not need to use them here either. 
On the other hand, our proof has to be more uniform than the proof in reverse mathematics.
For this purpose, we need some preliminary results about coproduct function spaces.

In the following we want to consider $X=\bigsqcup_I\CC(I,\IR^n)$ as a computable metric space. 
Here the coproduct is taken over all rational closed intervals $I=[a,b]\In\IR$ with $a,b\in\IQ$ and $a<b$,
represented as pairs $(a,b)\in\IQ^2$.
The following is easy to see and it is based on the standard construction of a metric for coproducts.

\begin{lemma}[Coproduct function space]
\label{lem:coproduct}
The space $X=\bigsqcup_I\CC(I,\IR^n)$ is a computable metric space endowed with the metric $d:X\times X\to\IR$, given by
\[d((f,I),(g,J)):=\left\{\begin{array}{ll}
\sup_{x\in I}\frac{||f(x)-g(x)||}{1+||f(x)-g(x)||} & \mbox{if $I=J$}\\
2 & \mbox{otherwise}
\end{array}\right.\]
\end{lemma}

A suitable computable dense subset can be constructed with the
help of rational polynomials. It is also clear that the injection into coproduct spaces is computable.

\begin{lemma}[Injection into coproduct spaces]
\label{lem:injection}
The canonical injection maps 
\begin{eqnarray*}
&\inj:\CC(J,\IR^n)\to\bigsqcup_I\CC(I,\IR^n),f\mapsto(f,J)
\end{eqnarray*}
are computable for every
$J=[a,b]$ with rational $a<b$.
\end{lemma}

Now we are prepared to prove the main result of this section. As usual, we denote by $B(x,r):=\{y\in X:d(x,y)<r\}$
the {\em open ball} in a metric space $(X,d)$ and by $\overline{A}$ the {\em closure} of a set $A\In X$.
On $\IR^n$ we use the {\em maximum metric}.

\begin{proposition}
\label{prop:IVP-WKL}
$\IVP\leqSW\WKL$.
\end{proposition}
This result even holds for a version of $\IVP$ whose outputs consist of functions $y:[a,b]\to\IR$ defined
on closed intervals $[a,b]$ with rational endpoints $a<b$.
\begin{proof}
Given $U\in\OO(\IR^{n+1})$, $f\in\CC(U,\IR^n)$ and $(x_0,y_0)\in U$ we can compute a rational $\delta>0$
such that 
\[K:=\overline{B((x_0,y_0),\delta)}=[x_0-\delta,x_0+\delta]\times\overline{B(y_0,\delta)}\In U.\]
This allows us to compute
\[M:=\max_{(s,t)\in K}||f(s,t)||+1\]
and $a,b\in\IQ$ with 
\[x_0-\frac{\delta}{M}\leq a<x_0<b\leq x_0+\frac{\delta}{M}.\]
If we can find a solution $y:[a,b]\to\IR^n$ of the initial value problem~(\ref{eq:IVP}), 
then also the restriction $y:(a,b)\to\IR^n$ to the open interval $(a,b)$ is a solution.
We let $I:=[a,b]$. By \cite[Theorem~2.19]{Tes12}
there exists a solution $y:I\to\IR^n$ of the initial value problem~(\ref{eq:IVP}).
Any such solution is in 
\[D:=\{y\in\CC(I,\IR^n):y(x_0)=y_0\mbox{ and }(\forall s,t\in I)\,||y(s)-y(t)||\leq M|s-t|\}\]
and any $y\in D$ satisfies $y(I)\In\overline{B(y_0,\delta)}$.
We consider the {\em Picard operator}
\begin{eqnarray}
T:D\to D,y\mapsto \left(x\mapsto y_0+\int_{x_0}^x f(t,y(t))\;\mathrm{d}t\right)
\label{eqn:Picard}
\end{eqnarray}
for this domain $D$. It is not too difficult to see that $T$ is well-defined, i.e., $T(D)\In D$.
Given the input data $(f,U,x_0,y_0)$, we can also compute a name of $T\in\CC(D,D)$, as integration is computable by~\cite[Theorem~6.4.1]{Wei00}.
The set 
\[A:=\{y\in D:T(y)=y\}\] 
of fixed points of $T$ contains exactly the solutions $y:I\to\IR^n$ of the 
initial value problem~(\ref{eq:IVP}) (see~\cite[Section~2.2]{Tes12}).
We can compute a name of $A\in\AA_-(D)$ as a closed set.
We claim that we can also compute a name of $D\in\KK_-(\CC(I,\IR^n))$ as a compact set.
Then it follows by~\cite[Proposition~5.5~(4)]{Pau16} that we can also compute a name of $A\in\KK_-(\CC(I,\IR^n))$
as a compact set. By Lemma~\ref{lem:coproduct} also the coproduct space $X=\bigsqcup_I\CC(I,\IR^n)$
is a computable metric space and the natural injection of $\CC(I,\IR^n)$ into $X$
is computable by Lemma~\ref{lem:injection}. Hence, it can be lifted to a computable
injection from $\KK_-(\CC(I,\IR^n))$ into $\KK_-(X)$ by~\cite[Proposition~5.5~(6)]{Pau16}.
Altogether, with Theorem~\ref{thm:K-WKL} this yields the reduction 
\[\IVP\leqSW\K_X\leqSW\WKL.\]

It remains to prove the claim on computability of $D$ as a compact subset of $\CC(I,\IR^n)$.
For technical simplicity, we describe the construction for the case $n=1$. The general case
can be treated similarly. 
Let $(q_i)_{i\in\IN}$ be a computable enumeration of the rational numbers in $I=[a,b]$.
We claim that the following map is a computable embedding relative to the input data
\[e:D\to[-1,1]^\IN,y\mapsto\left(\frac{1}{\delta}(y(q_i)-y_0)\right)_{i\in\IN}.\]
Firstly, $e$ is well-defined as $y\in D$ implies $y(I)\In\overline{B(y_0,\delta)}$,
and $e$ is obviously computable relative to the input data. 
Since two $y_1,y_2\in D$ with $y_1\not=y_2$ must differ on a rational input, 
it follows that $e$ is injective. Now we consider the set
\[C:=\left\{z\in[-1,1]^\IN:(\forall i,j)\;||z_i-z_j||\leq\frac{M}{\delta}\cdot|q_i-q_j|\right\}.\]
We claim that there is a function $g:C\to\CC(I,\IR^n)$ that is a left inverse to $e$ and computable relative
to the input data. Firstly, it is clear that $e(D)\In C$.
Given $z\in C$, there is a uniquely defined continuous function $y:I\to\IR^n$ given by
\[y(q_i):=y_0+\delta\cdot z_i\]
for all $i\in\IN$. If we set $g(z):=y$, then it is clear that $g$ is left inverse to $e$.
We still need to prove that $g$ is computable.
The definition of $y$ implies
\[||y(q_i)-y(q_j)||\leq M\cdot|q_i-q_j|\]
for all $i,j\in\IN$.
Hence, in order to evaluate $g(z)=y$ up to precision $2^{-k}$, it suffices to
find some $j\in\IN$ with $|x-q_j|<\frac{2^{-k}}{M}$. Then $||y(x)-y(q_j)||\leq 2^{-k}$ follows.
This also proves that $e$ is an embedding and $e$ and its partial inverse $e^{-1}$ are both
computable relative to the input data.

If we can now prove that $e(D)$ is co-c.e.\ closed in the compact metric space $[-1,1]^\IN$ relative
to the input data, then it follows that $D=g\circ e(D)$ is a computable point in $\KK_-(\CC(I,\IR^n))$ relative
to the input data by~\cite[Proposition~5.5]{Pau16}. Now, if $z\in[-1,1]^\IN\setminus e(D)$ then either $z\not\in C$, which
can be recognized or $z\in C\setminus e(D)$, which means $g(z)(x_0)\not=y_0$, which can also
be recognized. Altogether, this proves the claim.
\QED
\end{proof}

The proof that $D$ is computably compact could also be obtained with the help of a
suitable computable version of the Arzel\`a-Ascoli theorem. Instead of using the computable metric space $X$,
we could also directly prove $\IVP\leqSW\K_{[-1,1]^\IN}$.

As an immediate corollary of Proposition~\ref{prop:IVP-WKL} we obtain 
the following result, which is a version of Corollary~\ref{cor:main} for the non-maximal case.

\begin{corollary}
\label{cor:IVP-WKL}
Let $f:U\to\IR^n$ be computable with a c.e.\ open set $U\In\IR\times\IR^n$ and computable $(x_0,y_0)\in U$.
Then there exists a solution $y:I\to\IR^n$ of (\ref{eq:IVP}) with $I=[a,b]$ and rational $a<b$
such that $y$ is low as a point in $\CC(I,\IR^n)$. And given $(f,U,x_0,y_0)$, a solution $(y,I)$ can be found
uniformly in a non-deterministic way.
\end{corollary}

For $y$ to be {\em low} in $\CC(I,\IR^n)$ means that $y$ has a name $p\in\IN^\IN$ that is low, i.e., whose
Turing jump is computable relative to the halting problem.
We note that a low point $y\in\CC(I,\IR^n)$ is not the same thing as a low function $y:I\to\IR^n$
(see~\cite{Bra18} for this distinction).

\section{Reduction of $\IVP_{\max}$ to $\WKL$}
\label{sec:IVP-MAX-WKL}

In this section we want to strengthen Proposition~\ref{prop:IVP-WKL} in the sense that we
can even solve initial value problems for their maximal domains of existence with the help of $\WKL$.
For this strengthening we will actually use a refined version of Proposition~\ref{prop:IVP-WKL} repeatedly.
In fact, we will follow the proof idea of Gra{\c{c}}a, Zhong, and Buescu~\cite{GZB09}, which 
essentially is to apply $\IVP(f,U,x_0,y_0)$ repeatedly with values $x_0,y_0$ at the boundary of an already
existing solution, in order to extend the domain of the solution step by step. 

\begin{proposition}
\label{prop:IVP-MAX-WKL}
$\IVP_{\max}\leqSW\WKL$.
\end{proposition}
\begin{proof}
We first refine the proof of Proposition~\ref{prop:IVP-WKL} by making the choice of $\delta>0$ and $a,b\in\IQ$
more specific. Let $(f,U,x_0,y_0)$ be the input given to $\IVP$.
We can assume, without loss of generality, that there are $c_m\in\IR^{n+1}$ and $r_m>0$ such that
\[U=\bigcup_{m\in\IN}B(c_m,r_m)\mbox{ with }\overline{B(c_m,r_m)}\In U\] 
for all $m\in\IN$. For each pair $\langle m,k\rangle\in\IN$ at least one of the conditions
\[||c_m-(x_0,y_0)||<r_m-2^{-k}\mbox{ or }||c_m-(x_0,y_0)||> r_m-2^{-k+1}\]
has to hold, and we check for each pair, which condition we can verify first, until 
we have found a pair $\langle m',k'\rangle$ for which we can recognize the first condition first.
Such a pair needs to exist, as $k'$ can be made sufficiently large. This guarantees that the 
$\langle m',k'\rangle\in\IN$ that we have found is smaller than or equal to the first $\langle m,k\rangle$ with
\begin{eqnarray}
||c_m-(x_0,y_0)||\leq r_m-2^{-k+1}.
\label{eqn:IVP-MAX-WKL1}
\end{eqnarray}
We set $\delta:=2^{-k'}$ and we choose $M$ and $a,b\in\IQ$ as in the proof of Proposition~\ref{prop:IVP-WKL},
but with the additional property that
\begin{eqnarray}
&x_0-a>\frac{1}{2}\cdot\frac{\delta}{M}\mbox{ and }b-x_0>\frac{1}{2}\cdot\frac{\delta}{M}.
\label{eqn:IVP-MAX-WKL2}
\end{eqnarray}
We obtain $\overline{B((x_0,y_0),\delta)}\In B(c_{m'},r_{m'})$ and hence $M\leq M_{m'}$
for
$M_{m'}:=\max_{z\in\overline{B(c_{m'},r_{m'})}}||f(z)||+1$.

We consider a version of $\IVP$ that produces an output $y:[a,b]\to\IR^n$ with the above additional
requirements~(\ref{eqn:IVP-MAX-WKL1}) and (\ref{eqn:IVP-MAX-WKL2}). 
The remainder of the proof of Proposition~\ref{prop:IVP-WKL} still shows $\IVP\leqSW\WKL$ for this version of $\IVP$. 

Now we use this specific version of $\IVP$ in an infinite loop inductively as follows.
Given $(f,U,x_0,y_0)$ as input to $\IVP_{\max}$, 
we choose $a_0:=b_0:=x_0$, $y_{a_0}(a_0):=y_{b_0}(b_0):=y_0$. 
Then we determine inductively in a loop for all $i\in\IN$
\begin{itemize}
\item $(y_{a_{i+1}},[a_{i+1},.])\in\IVP(f,U,a_i,y_{a_i}(a_i))$,
\item $(y_{b_{i+1}},[.,b_{i+1}])\in\IVP(f,U,b_i,y_{b_i}(b_i))$.
\end{itemize}
The infinite loop that determines these values can be realized with the help of $(\IVP\times\IVP)^\infty$.
The final result of this computation is $(y,(a,b))$ with $a:=\inf_{i\in\IN} a_i$, $b:=\sup_{i\in\IN}b_i$ and
$y:(a,b)\to\IR^n$ with
\[y(x):=\left\{\begin{array}{ll}
y_{a_{i+1}}(x) & \mbox{if $a_{i+1}\leq x\leq a_i$}\\
y_{b_{i+1}}(x) & \mbox{if $b_i\leq x\leq b_{i+1}$}
\end{array}\right..\]
The set $(a,b)\in\OO(\IR)$ and the function $y\in\CC((a,b),\IR^n)$ 
can be computed relative to the previously determined objects, as
$y_{a_{i+1}}(a_i)=y_{a_i}(a_i)$ and $y_{b_{i+1}}(b_i)=y_{b_i}(b_i)$.
We claim that $y$ is a maximal solution of the initial value problem~(\ref{eq:IVP}).
With the help of Propositions~\ref{prop:IVP-WKL} and \ref{prop:WKL} this proves
\[\IVP_{\max}\leqSW(\IVP\times\IVP)^\infty\leqSW\WKL^\infty\equivSW\WKL.\]
We still need to prove the claim. 
Let us assume that the solution ${y:(a,b)\to\IR^n}$ is not maximal and that there
is some $b'>b$ such that $y$ can be extended to a solution $y:(a,b')\to\IR^n$.
Then $(b,y(b))\in U$ and there are $m,k,j\in\IN$ with 
\begin{eqnarray}
&B((b_j,y(b_j)),2^{-k+1})\In B((b,y(b)),2^{-k+2})\In B(c_m,r_m)\mbox{ and }
\label{eqn:IVP-MAX-WKL3}\\
&b_j+\frac{1}{2}\cdot\min_{\langle m',k'\rangle\leq\langle m,k\rangle}\frac{2^{-k'}}{M_{m'}}>b,
\label{eqn:IVP-MAX-WKL4}
\end{eqnarray}
since $\lim_{i\to\infty}b_i=b$. If we apply~(\ref{eqn:IVP-MAX-WKL1}) and (\ref{eqn:IVP-MAX-WKL2}) 
to the pair $(b_j,y(b_j))$ (in place of $(x_0,y_0)$), 
then we obtain that the $\langle m',k'\rangle$ chosen by the algorithm for $\IVP$
on input $(f,U,b_j,y(b_j))$ satisfies $\langle m',k'\rangle\leq\langle m,k\rangle$ by~(\ref{eqn:IVP-MAX-WKL1}) and
(\ref{eqn:IVP-MAX-WKL3})
and hence we obtain with $\delta=2^{-k'}$
\begin{eqnarray*}
&b_{j+1}> b_j+\frac{1}{2}\cdot\frac{\delta}{M_{m'}}\geq b_j+\frac{1}{2}\cdot\min_{\langle m',k'\rangle\leq\langle m,k\rangle}\frac{2^{-k'}}{M_{m'}}>b
\end{eqnarray*}
by~(\ref{eqn:IVP-MAX-WKL2}) and (\ref{eqn:IVP-MAX-WKL4}).
But this is a contradiction to $b=\sup_{i\in\IN}b_i$. Hence, there cannot be any extension
$y:(a,b')\to\IR^n$ of the solution $y:(a,b)\to\IR^n$ with $b'>b$.
Likewise, it follows that there cannot be any extension $y:(a',b)\to\IR^n$ with $a'<a$.
\QED
\end{proof}

Theorem~\ref{thm:Collins} and Corollaries~\ref{cor:main} and \ref{cor:finite} are immediate consequences
of Proposition~\ref{prop:IVP-MAX-WKL}.

\section{Reduction of $\WKL$ to $\IVP$}
\label{sec:WKL-IVP}

In this section we will translate the proof idea of Aberth~\cite{Abe71} into a proof of $\WKL\leqSW\IVP$.
This idea has also been used by Simpson~\cite[Theorem IV.8.1]{Sim09}.
However, the uniform version of this proof needs a little modification, which ensures
that the required information can be reconstructed from a solution of the constructed
initial value problem on some arbitrary small domain.

\begin{proposition}
\label{prop:WKL-IVP}
$\WKL\leqSW\IVP$.
\end{proposition}
\begin{proof}
By Proposition~\ref{prop:WKL} it suffices to prove $\widehat{\LLPO}\leqSW\IVP$ 
and for this purpose we first describe a gadget that
does the job for a single instance of $\LLPO$.
We can assume that
$\LLPO:\In\IN^\IN\mto\{0,1\}$ is defined for all $p\in\IN^\IN$ with $\{0,1\}\not\In\range(p)$ and all $i\in\{0,1\}$ by
\[i\in\LLPO(p)\iff i\not\in\range(p).\]
Given $p\in\IN^\IN$ we construct a continuous function $g_p:\IR^2\to\IR$ such that 
all solutions $y:[0,4]\to\IR$ of the 
corresponding initial value problem $y'(x)=g_p(x,y(x))$ with $y(0)=0$ 
satisfy $y(4)=0$ and
\begin{eqnarray}
&y(2)>-1\TO 0\in\LLPO(p)\label{eqn:LLPO-0}\\
&y(2)<1\TO  1\in\LLPO(p).\label{eqn:LLPO-1}
\end{eqnarray}
The idea is to construct the function $g_p$ such that the unique solutions are sufficiently positive or negative
for $\LLPO(p)=\{0\}$ and $\LLPO(p)=\{1\}$, respectively, and such that there are positive and negative
solutions for $\LLPO(p)=\{0,1\}$. Figure~\ref{fig:IVP} illustrates the situation.

\begin{figure}[htb]
\vspace{-0.5cm}
\includegraphics[width=4cm]{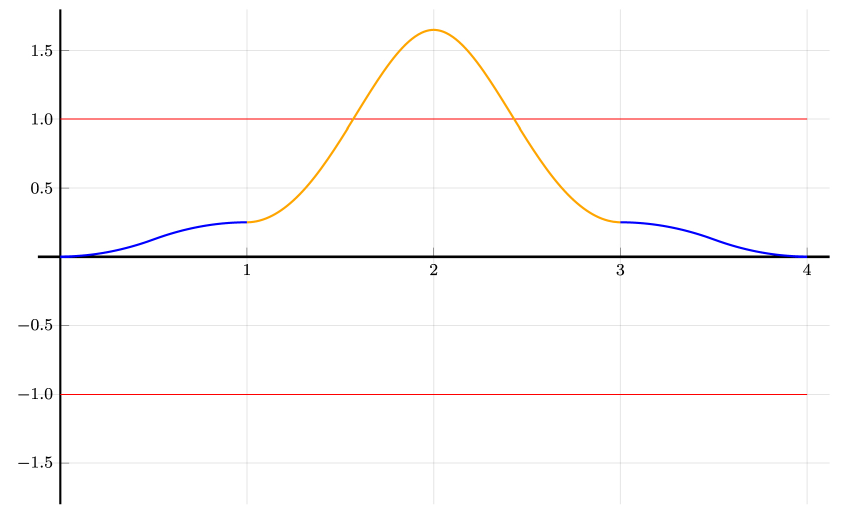}
\includegraphics[width=4cm]{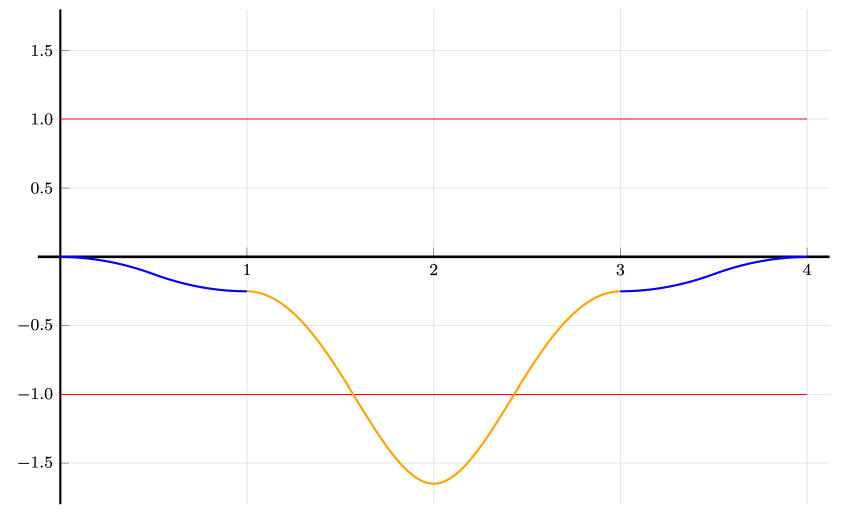}
\includegraphics[width=4cm]{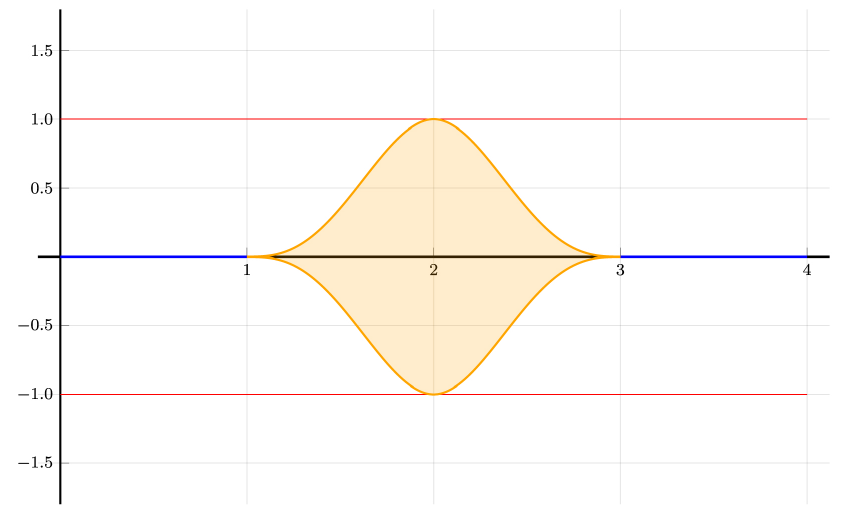}
\begin{center}
$\LLPO(p)=\{0\}$\hspace{2cm}$\LLPO(p)=\{1\}$\hspace{2cm}$\LLPO(p)=\{0,1\}$
\end{center}
\caption{Solutions $y$ of the initial value problem $y'(x)=g_p(x,y(x))$ with $y(0)=0$ .}
\label{fig:IVP}
\end{figure}

The function $g_p:\IR^2\to\IR$ is defined piecewise by 
\[g_p(x,y):=\left\{\begin{array}{ll}
h_p(x) & \mbox{if $x\in[0,1]$}\\
s(x-1,y) & \mbox{if $x\in[1,2]$}\\
-s(x-2,y) &\mbox{if $x\in[2,3]$}\\
-h_p(x-3) & \mbox{if $x\in[3,4]$}\\
0 & \mbox{otherwise}
\end{array}\right.,\]
where $h_p$ governs the blue parts and $s$ the orange parts of the solutions in Figure~\ref{fig:IVP}.
The function $h_p:[0,1]\to\IR$ is given by
\[h_p(x):=\max(0,1-|2x-1|)\cdot(T(p,1)-T(p,0))\]
with
\[T(p,i):=\left\{\begin{array}{ll}2^{-\min\{j\in\IN:p(j)=i\}}&\mbox{if $i\in\range(p)$}\\0&\mbox{otherwise}\end{array}\right..\]
That is $h_p$ is a triangle function with a peak at $\frac{1}{2}$ and value $2^{-k}$ or $-2^{-k}$, respectively,
depending on whether $1$ or $0$ first appears at position $k$ in $p$, respectively, and hence
$\LLPO(p)=\{0\}$ or $\LLPO(p)=\{1\}$, respectively. If neither of $0,1$ appears in $p$, then
$h_p$ is constantly $0$. 
The function $s:[0,1]\times\IR\to\IR$ is given by
\begin{center}
$s(x,y):=9x(1-x)\cdot\sign(y)|y|^\frac{1}{3}$.
\end{center}
The initial value problem $y'(x)=s(x,y(x))$ with $y(0)=y_0$ has the solution
\begin{center}
$y(x)=\sign(y_0)\left(x^2(3-2x)+|y_0|^\frac{2}{3}\right)^\frac{3}{2}$
\end{center}
for $y_0\not=0$, and for $y_0=0$ and all $c\in[0,1]$ one has the solutions
\begin{center}
$y(x)=\left\{\begin{array}{ll}
0 & \mbox{if $x\leq c$}\\
\pm\left(x^2(3-2x)-c^2(3-2c)\right)^\frac{3}{2} & \mbox{if $c\leq x\leq1$}
\end{array}\right..$
\end{center}
As explained in \cite{Abe71}, these are all possible solutions of the given initial value problem for $s$.
The solutions $y$ for $y_0>0$ satisfy $y(1)>1$ and the solutions for $y_0<0$ satisfy $y(1)<-1$
and hence the solutions $y$ for the initial value problem of $g_p$ satisfy the properties
given in (\ref{eqn:LLPO-0}) and (\ref{eqn:LLPO-1}).

In a second step we combine a countable number of the gadgets for a single instance 
of $\LLPO$ in a single initial value problem in order to reduce the problem $\widehat{\LLPO}$ to $\IVP$.
Given an input $p=\langle p_0,p_1,p_2,...\rangle$ of $\widehat{\LLPO}$, we can compute the function 
$f:[-1,1]\times\IR\to\IR$ with
\[
f(x,y):=\left\{\begin{array}{ll}
\sum_{m=\langle k,i\rangle\in\IN} 2^{-(m+3)}g_{p_i}\left(2^{m+3}(x+2^{-m}),2^{2(m+3)}y\right) & \mbox{if $x\leq0$}\\
f(-x,y) & \mbox{otherwise}
\end{array}\right..\]
We note that the construction ensures that the information on any value of $\LLPO(p_i)$ is included for infinitely many $m$ in $f$,
which is necessary as a solution ${y:I\to\IR}$ of the initial value problem $y'(x)=f(x,y(x))$ with $y(0)=0$
might only be known on some small interval $I=(a,b)$ around $0$. 
We claim that any such solution $y$ satisfies 
\begin{eqnarray}
&y(-2^{-m}+2^{-(m+2)})>-2^{-(m+3)}\TO 0\in\LLPO(p_i)\label{eqn:LLPOm-0}\\
&y(-2^{-m}+2^{-(m+2)})<2^{-(m+3)}\TO  1\in\LLPO(p_i)\label{eqn:LLPOm-1}
\end{eqnarray}
for all $m=\langle k,i\rangle\in\IN$ with $-2^{-m}+2^{-(m+2)}\in I$.
This enables us to compute a value $q\in\widehat{\LLPO}(p)$ from $\IVP(f,(-1,1)\times\IR,0,0)$
and hence we obtain 
\[\WKL\equivSW\widehat{\LLPO}\leqSW\IVP.\]
It still remains to show how the claim follows from the implications 
given in (\ref{eqn:LLPO-0}) and (\ref{eqn:LLPO-1}).
We consider $m=\langle k,i\rangle\in\IN$ with $-2^{-m}+2^{-(m+2)}\in I$
and solutions $y$ of $y'(x)=f(x,y(x))$ with $y(0)=0$ and solutions $\hat{y}$ of $\hat{y}'(\hat{x})=g_{p_i}(\hat{x},\hat{y}(\hat{x}))$ with $\hat{y}(0)=0$ on the intervals $[-2^{-m},-2^{-(m+1)}]$ and $[0,4]$, respectively.
The transformation $\hat{x}=2^{m+3}(x+2^{-m})$ maps the interval $[-2^{-m},-2^{-(m+1)}]$
onto the interval $[0,4]$. Together with the transformation $\hat{y}=2^{2(m+3)}y$ we obtain
\begin{eqnarray*}
\frac{{\rm d}y}{{\rm d}x} &=& 2^{-2(m+3)}\cdot\frac{{\rm d}\hat{y}}{{\rm d}y}\cdot\frac{{\rm d}y}{{\rm d}x}\cdot\frac{{\rm d}x}{{\rm d}\hat{x}}\cdot2^{m+3}=2^{-(m+3)}\cdot \frac{{\rm d}\hat{y}}{{\rm d}\hat{x}}=2^{-(m+3)}\cdot g_{p_i}(\hat{x},\hat{y}(\hat{x}))
\end{eqnarray*}
and hence
\[y(-2^{-m}+2^{-(m+2)})=2^{-(m+3)}\hat{y}(2).\]
This shows that (\ref{eqn:LLPO-0}) and (\ref{eqn:LLPO-1}) imply (\ref{eqn:LLPOm-0}) and (\ref{eqn:LLPOm-1}), respectively.
\QED
\end{proof}

We note that this reduction requires $\IVP$ only in the special case $x_0=y_0=0$.

\begin{credits}
\subsubsection{\ackname} 
The first author was funded by the German Research Foundation (DFG, Deutsche Forschungsgemeinschaft) -- project number 554999067 and by the National Research Foundation of South Africa (NRF) -- grant number 151597.
\end{credits}

%
% ---- Bibliography ----
%
% BibTeX users should specify bibliography style 'splncs04'.
% References will then be sorted and formatted in the correct style.
%
\bibliographystyle{splncs04}
\bibliography{C:/Users/\input{"|kpsewhich -var-value USERNAME"}/Documents/Spaces/Research/Bibliography/lit}

\begin{thebibliography}{10}
\providecommand{\url}[1]{\texttt{#1}}
\providecommand{\urlprefix}{URL }
\providecommand{\doi}[1]{https://doi.org/#1}

\bibitem{Abe71}
Aberth, O.: The failure in computable analysis of a classical existence theorem
  for differential equations. Proceedings of the American Mathematical Society
  \textbf{30},  151--156 (1971)

\bibitem{Bra18}
Brattka, V.: A {G}alois connection between {T}uring jumps and limits. Logical
  Methods in Computer Science  \textbf{14}(3:13),  1--37 (Aug 2018).
  \doi{10.23638/LMCS-14(3:13)2018}, \url{https://lmcs.episciences.org/4794}

\bibitem{Bra24}
Brattka, V.: Loops, inverse limits and non-determinism (2024), unpublished
  notes

\bibitem{BBP12}
Brattka, V., de~Brecht, M., Pauly, A.: Closed choice and a uniform low basis
  theorem. Annals of Pure and Applied Logic  \textbf{163},  986--1008 (2012).
  \doi{10.1016/j.apal.2011.12.020},
  \url{http://dx.doi.org/10.1016/j.apal.2011.12.020}

\bibitem{BG11}
Brattka, V., Gherardi, G.: Weihrauch degrees, omniscience principles and weak
  computability. The Journal of Symbolic Logic  \textbf{76}(1),  143--176
  (2011). \doi{10.2178/jsl/1294170993},
  \url{http://dx.doi.org/10.2178/jsl/1294170993}

\bibitem{BGP21}
Brattka, V., Gherardi, G., Pauly, A.: Weihrauch complexity in computable
  analysis. In: Brattka, V., Hertling, P. (eds.) Handbook of Computability and
  Complexity in Analysis, pp. 367--417. Theory and Applications of
  Computability, Springer, Cham (2021). \doi{10.1007/978-3-030-59234-9\_11},
  \url{https://doi.org/10.1007/978-3-030-59234-9\_11}

\bibitem{BH21}
Brattka, V., Hertling, P. (eds.): Handbook of Computability and Complexity in
  Analysis. Theory and Applications of Computability, Springer, Cham (2021).
  \doi{10.1007/978-3-030-59234-9},
  \url{https://doi.org/10.1007/978-3-030-59234-9}

\bibitem{BLRMP19}
Brattka, V., Le~Roux, S., Miller, J.S., Pauly, A.: Connected choice and the
  {B}rouwer fixed point theorem. Journal of Mathematical Logic  \textbf{19}(1),
   1--46 (2019). \doi{10.1142/S0219061319500041},
  \url{https://doi.org/10.1142/S0219061319500041}

\bibitem{CG09}
Collins, P., Gra{\c{c}}a, D.S.: Effective computability of solutions of
  differential inclusions: the ten thousand monkeys approach. Journal of
  Universal Computer Science  \textbf{15}(6),  1162--1185 (2009).
  \doi{10.3217/jucs-015-06-1206},
  \url{http://dx.doi.org/10.3217/jucs-015-06-1206}

\bibitem{GM09}
Gherardi, G., Marcone, A.: How incomputable is the separable {H}ahn-{B}anach
  theorem? Notre Dame Journal of Formal Logic  \textbf{50}(4),  393--425
  (2009). \doi{10.1215/00294527-2009-018},
  \url{http://dx.doi.org/10.1215/00294527-2009-018}

\bibitem{GZB09}
Gra{\c{c}}a, D.S., Zhong, N., Buescu, J.: Computability, noncomputability and
  undecidability of maximal intervals of {IVP}s. Transactions of the American
  Mathematical Society  \textbf{361}(6),  2913--2927 (2009).
  \doi{10.1090/S0002-9947-09-04929-0},
  \url{http://dx.doi.org/10.1090/S0002-9947-09-04929-0}

\bibitem{GZ21}
Gra{\c{c}}a, D.S., Zhong, N.: Computability of differential equations. In:
  Brattka, V., Hertling, P. (eds.) Handbook of Computability and Complexity in
  Analysis, pp. 71--99. Theory and Applications of Computability, Springer,
  Cham (2021). \doi{10.1007/978-3-030-59234-9\_3},
  \url{https://doi.org/10.1007/978-3-030-59234-9\_3}

\bibitem{Hau85}
Hauck, J.: {E}in {K}riterium f\"{u}r die konstruktive {L}\"{o}sbarkeit der
  {D}ifferentialgleichung $y'=f(x,y)$. Zeitschrift f\"ur Mathematische Logik
  und Grundlagen der Mathematik  \textbf{31},  357--362 (1985)

\bibitem{LRP15a}
Le~Roux, S., Pauly, A.: Finite choice, convex choice and finding roots. Logical
  Methods in Computer Science  \textbf{11}(4),  4:6, 31 (2015).
  \doi{10.2168/LMCS-11(4:6)2015},
  \url{http://dx.doi.org/10.2168/LMCS-11(4:6)2015}

\bibitem{Pau16}
Pauly, A.: On the topological aspects of the theory of represented spaces.
  Computability  \textbf{5}(2),  159--180 (2016). \doi{10.3233/COM-150049},
  \url{http://dx.doi.org/10.3233/COM-150049}

\bibitem{PR79}
Pour-El, M.B., Richards, J.I.: A computable ordinary differential equation
  which possesses no computable solution. Annals Math.\ Logic  \textbf{17},
  61--90 (1979)

\bibitem{Ruo96}
Ruohonen, K.: An effective {C}auchy-{P}eano existence theorem for unique
  solutions. International Journal of Foundations of Computer Science
  \textbf{7}(2),  151--160 (1996). \doi{10.1142/S0129054196000129},
  \url{https://doi.org/10.1142/S0129054196000129}

\bibitem{Sch21}
Schr{\"o}der, M.: Admissibly represented spaces and {Qcb}-spaces. In: Brattka,
  V., Hertling, P. (eds.) Handbook of Computability and Complexity in Analysis,
  pp. 305--346. Theory and Applications of Computability, Springer, Cham
  (2021). \doi{10.1007/978-3-030-59234-9\_9},
  \url{https://doi.org/10.1007/978-3-030-59234-9\_9}

\bibitem{Sim84}
Simpson, S.: Which set existence axioms are needed to prove the
  {C}auchy/{P}eano theorem for ordinary differential equations? The Journal of
  Symbolic Logic  \textbf{49},  783--802 (1984). \doi{10.2307/2274131},
  \url{https://doi.org/10.2307/2274131}

\bibitem{Sim09}
Simpson, S.G.: Subsystems of Second Order Arithmetic. Perspectives in Logic,
  Cambridge University Press, 2nd edn. (2009)

\bibitem{Smi24}
Smischliaew, H.A.: Berechenbarkeitseigenschaften von {D}ifferentialgleichungen.
  Master's thesis, Fakult{\"a}t f{\"u}r Informatik, Universit{\"a}t der
  Bundeswehr M{\"u}nchen, Neubiberg (2024)

\bibitem{Tes12}
Teschl, G.: Ordinary differential equations and dynamical systems, Graduate
  Studies in Mathematics, vol.~140. American Mathematical Society, Providence,
  RI (2012). \doi{10.1090/gsm/140}, \url{https://doi.org/10.1090/gsm/140}

\bibitem{Wei00}
Weihrauch, K.: Computable Analysis. Springer, Berlin (2000)

\end{thebibliography}

\end{document}